# A facile direct device transfer of monolayer MoS$_2$ towards improvement in transistor performances

Sameer Kumar Mallik,[1,2] Roshan Padhan,[1,2] Suman Roy,[1,2] Mousam Charan Sahu,[1,2] Sandhyarani Sahoo,[1,2] Satyaprakash Sahoo[1,2,*]

[1]*Laboratory for Low Dimensional Materials, Institute of Physics, Bhubaneswar-751005, India*
[2]*Homi Bhabha National Institute, Training School Complex, Anushakti Nagar, Mumbai, 400094, India*

*Corresponding Author email: sahoo@iopb.res.in

## Abstract

Transfer techniques based on two dimensional (2D) materials and devices offer immense potential towards their industrial integration with the existing silicon based electronics. To achieve high quality devices, there is an urgent requirement for the etching-free, and clean transfer that retain original semiconducting properties of layered channel materials. In parallel, transfer of metal electrode arrays on the 2D semiconductors also attract attention towards large-scale integration for commercial applications. Here, we demonstrate a facile PMMA-assisted etching-free one-step approach to transfer both 2D channels and metal electrodes without damaging the contact region. The direct device transfer (DDT) technique enables residue-free monolayer MoS$_2$ as channel material towards achieving doping-free intrinsic transistors with enhanced performances. The crystalline quality, strain relaxation, and interfacial coupling effects are studied using Raman and photoluminescence spectra with spatial mapping. Post device transfer, a reduced pinning effect is observed by the effective modulation of gate tunable drain currents in MoS$_2$ transistors at room temperature. Furthermore, the extracted Schottky barrier heights, temperature dependence of threshold voltage shifts, hysteresis evolution, and mobility enhancements validates the improved transistor performances in transferred devices. The proposed DDT method can be utilized to directly transfer array of devices of 2D materials and heterostructures skipping various cumbersome steps in between and hence could offer high performance reliable electronic applications.

**Introduction**

Atomically thin two dimensional (2D) transition metal dichalcogenides (TMDCs) exhibit wide range of semiconducting properties, making them potential candidates for low-power electronic device applications in the limit of atomic thickness.[1] Owing to their van der Waals (vdW) bonded layered structure, wide band gap variability,[2] high mechanical flexibility,[3] and controllable opto-electronic responses,[4,5] the TMDCs could serve as potential replacement beyond silicon in front end of line (FEOL) technologies.[6–8] The recent industrial implementation of 2D materials in alternative back end of line (BEOL) integration[9,10] and Internet of Things (IoT)[11,12] based embedded sensors spurs the necessity for rapid production, and transfer of high quality channel materials with robust device fabrication. Chemical vapour deposition (CVD) is one of the most explored synthesis route to produce large-scale high-quality monolayer 2D semiconductors such as molybdenum disulfide ($MoS_2$).[13] The growth of such crystals at high temperatures (> 650 °C) often delivers inevitable effects such as charge/ion doping, contaminations on the growth substrate which could hinder the seamless integration of low-power electronics.[14] For example, recently, we have reported $Na^+$ ion doping in gate dielectric via salt-assisted CVD method that resulted large hysteresis in the transfer characteristics of monolayer $MoS_2$ based field-effect transistors (FETs).[15] The controlled interfacial engineering in such cases have potential advantages in multi-bit memory and neuromorphic computing applications. These CVD grown samples could be further explored to realize conventional transistor properties or flexible electronic applications by transferring to application specific target substrates.[16]

Considerable efforts have been made to transfer atomically thin 2D layers from growth substrates to desired fresh substrates, which include etchant-based PMMA-assisted wet transfer method, water soluble layer transfer method, polymer free transfer methods, metal assisted transfer methods, and PDMS-based dry transfer method etc.[17,18] However, conventional chemical etchants such as NaOH, and KOH demarcate the film quality due to their high corrosivity and chemical residues which significantly degrade the device performances.[19,20] Recently, Zhang et. al.[21] demonstrate a capillary force driven etchant free transfer method with several different TMDCs/substrate combinations which are free from etchant induced damage. However, lack of mechanical support results in unwanted bubbles and wrinkles making unfavourable conditions to achieve desired electrical characteristics.[21] In another approach, a double support layer PMMA/rosin is utilized for facile transfer of large area 2D materials to fabricate high performance electronic devices.[22] Similarly, Desai et. al. demonstrates Au-mediated exfoliation and etching-based transfer of monolayer $MoS_2$ (lateral dimension > 500 $\mu m$) for efficient opto-electronic performances.[23] Additionally, dry deterministic transfer methods using viscoelastic stamping such as PDMS and optional polypropylene carbonate (PPC) layer have recently been explored to fabricate high quality vdW heterostructure devices.[24,25]

Considering the potential applications of 2D TMDCs in emerging technologies, it is crucial to ensure the exceptional quality and faithfulness of the transferred films to safeguard the integrity of their material characteristics. As many transfer methods are susceptible to process-related vulnerabilities, including the presence of trapped bubbles, residual polymers, as well as cracks and wrinkles.[18] These characteristics have the potential to undermine the performance of the devices. The impact of commonly used PMMA-assisted methods, such as doping of 2D channels with polymer residues, strain effects etc. can be non-destructively characterized by using Raman and Photoluminescence (PL) spectroscopy.[26,27] For example, inhomogeneous or uncontrollable strain has adverse consequences on PL and optical applications,[28] while wrinkles, cracks or residual polymers can significantly impact device conductivity and carrier mobility.[29,30] To further improve the film quality, carrier injection from metal to channel material, and reduce interfacial roughness, alternative approaches with metal electrode transfer methods have been recently demonstrated.[31–33] The delamination of printed contacts are performed using graphene, hBN or PVA-assisted state-of-art techniques and transferred to the desired 2D films which allow the formation of vdW contacts with low contact resistances.[31] However, the transfer techniques that have discussed so far are limited to only layer or metal transfer and a method involving whole device transfer is still lacking. Moreover, efficient stacking engineering and monolithic integration of 2D materials in FEOL and BEOL technologies in CMOS scaling require further development of clean, damage-free, and cost-effective transfer of both channel materials and printed devices.

In this article, we demonstrate a facile one-step PMMA-assisted etchant-free technique that are employed to simultaneously transfer monolayer $MoS_2$ layers with metal contacts. The delamination of silver (Ag) metal electrodes with atomically thin monolayer channel $MoS_2$ from the growth substrate can be facilitated by several optimizations such as PMMA thickness, spin-coat parameters, baking time, and temperature etc. Furthermore, Raman and PL studies reveals the release of strain and coupling effects before and after transfer. The newly developed direct device transfer (DDT) method yields few notable alterations in electrical characteristics, especially the hysteresis, and transfer curves along with significant enhancement of transistor performances such as mobility. Additionally, the temperature dependent transport is carried out in $MoS_2$ field-effect transistors (FETs) to study the carrier dynamics and trapping mechanisms before and after DDT. Our technique unleashes future directions to efficiently transfer large array of 2D materials based devices and sensors for flexible CMOS integrations.

## Results and Discussion

The monolayer MoS$_2$ triangular domains are synthesized using a salt-assisted CVD method; the details of which are reported elsewhere.[15] The number of layers and crystal qualities are confirmed by considering the frequencies and line widths of Raman and PL spectra.[13,34] Devices based on two terminal geometries are fabricated on the CVD grown monolayer MoS$_2$ by using photolithography where the heavily doped Si$^{++}$ can serve as back gate for FET operations. As discussed, these devices (Ag electrode + MoS$_2$ channel) in whole can be transferred to a fresh target substrate without any damage. The schematics of such a transfer technique and the proposed DDT method are demonstrated in Fig. 1. The experiment is carried out on fabricated MoS$_2$ based FETs on SiO$_2$/Si$^{++}$ substrate with Ag as source/drain electrodes, as shown in the optical micrograph of Fig. 1 (top left). In the first step, a comparatively thick layer of PMMA is spin coated on such 1×1 cm$^2$ growth substrate containing MoS$_2$ FETs with a speed of 500 rpm for 120 s. The PMMA coated substrate is baked at 75 °C for 300 s and then, one edge of the substrate is scratched by using a blade followed by the substrate is immersed in a beaker of DI water for 180 s. This allows the water penetration through the scratched region between the PMMA film and the growth substrate which partially helps in delaminating the PMMA/device stack. It is worth mentioning that in some cases during salt-assisted CVD growth, a water soluble layer (Na$_2$S/Na$_2$SO$_4$) is generally associated with the growth substrate and could be present underneath MoS$_2$ layers, as reported previously.[17] The dissolution of such layers could be an additional factor to remove any strong adhesion of the PMMA stack towards the growth substrate. As shown in Fig. 1, the PMMA/device stack is carefully peeled out from the substrate in a single attempt from the scratched area. Extreme precautions are taken care during the peeling off stacked layers to get rid of any strain-induced damage or nanofractures. Nonetheless, PMMA serves as a robust supporting layer with good flexibility and adhesive contact to atomically thin monolayers as compared to other polymers.[18]

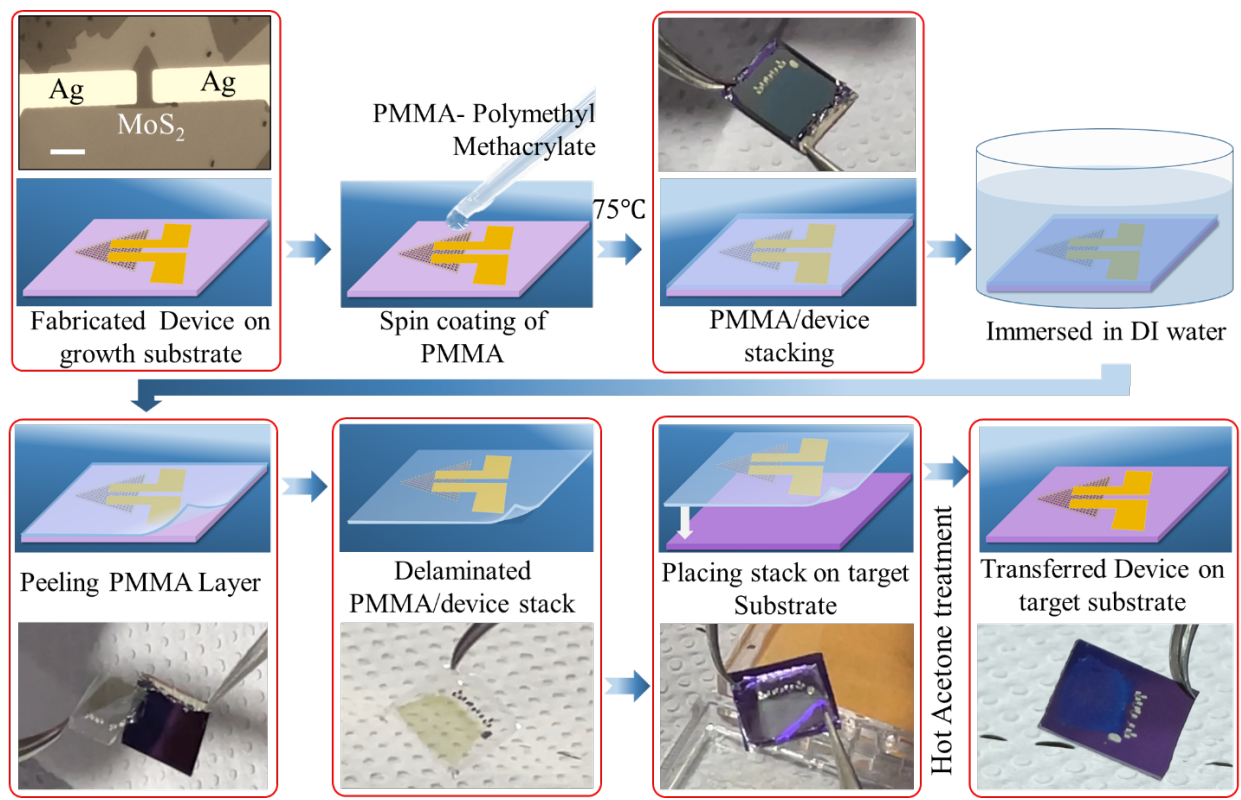

**Figure 1** *Realtime optical images and corresponding schematic representation of various steps during PMMA-assisted transfer of monolayer MoS$_2$ based Devices from growth substrate to target substrate. The scale bar is 10 μm.*

The delaminated array of devices with monolayer MoS$_2$ with PMMA adhesive layer are captured using camera images as represented in Fig. 1. Afterward, the PMMA/device stack are transferred onto a clean target substrate as demonstrated schematically. It may be noted that, the PMMA-assisted transfer induces large amount of polymer residues on the 2D surfaces.[35] Previously, these residues has been cleaned by various post-transfer methods such as annealing,[36] plasma treatment,[37] mechanical cleaning via conducting mode AFM,[38] and modified RCA cleaning.[39] However, we demonstrate a much simpler process to achieve residue-free surfaces post

PMMA-assisted DDT transfer by cleaning in hot acetone. The post transferred substrate is then dipped in hot acetone (45 °C) for 15 mins to not only dissolve the sacrificial PMMA layer, but also cleaning the polymer residues over monolayer MoS$_2$ channels. Hot acetone is also beneficial to avoid any deposition of the moisture after successfully transfer of the devices. The image of the transferred device to a clean SiO$_2$/Si substrate is shown in the bottom right of the Fig 1. The structural induced changes associated with strain and coupling are further examined by Raman and PL analysis before and after DDT. The electrical characterizations of as-fabricated and transferred devices are validated using temperature dependent transport measurements. It is important to highlight that we have repeated our DDT process for several devices, and the smooth transfer of whole devices are quite consistent.

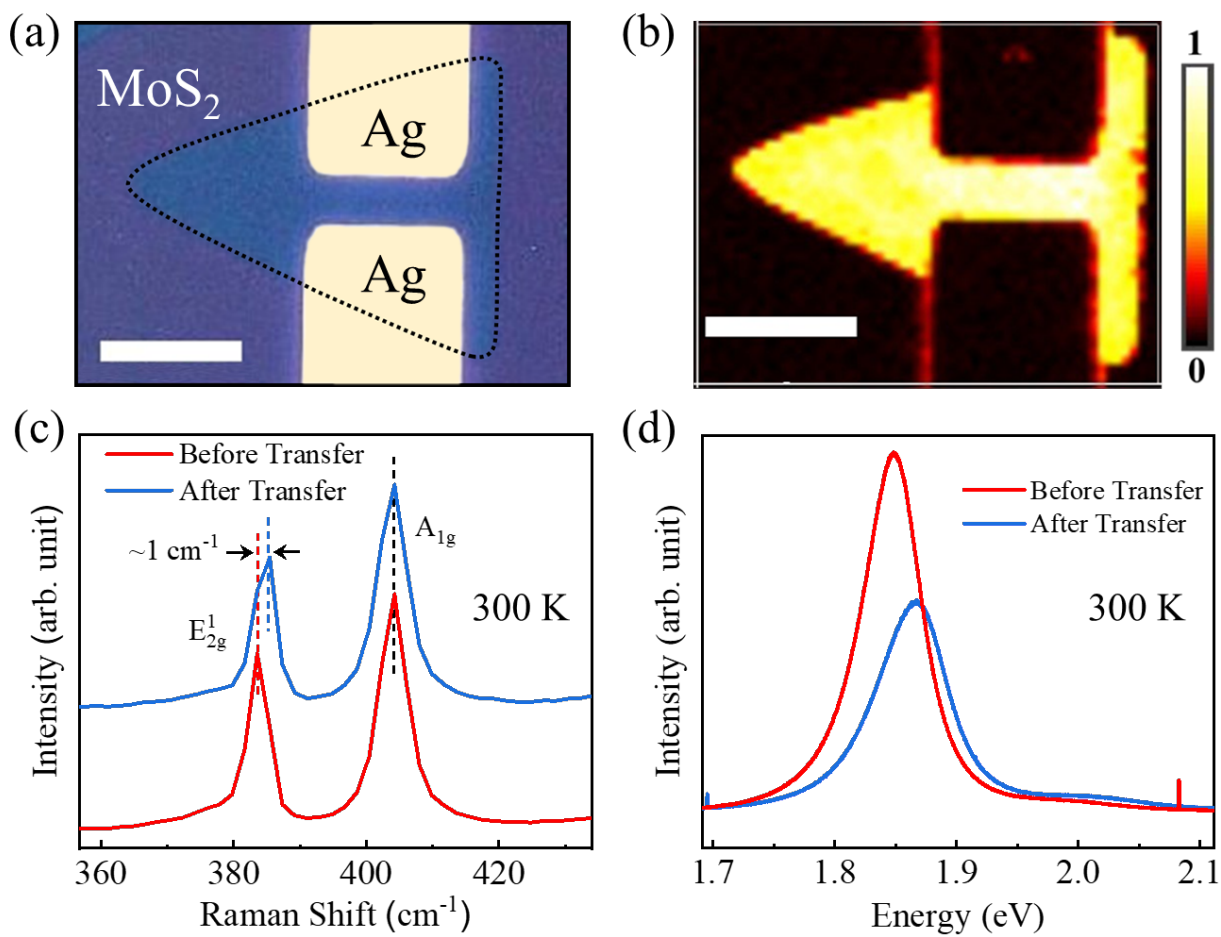

**Figure 2** *(a) Optical microscopy image of the transferred device (Ag/MoS₂) on a target SiO₂/Si⁺⁺ substrate. (b) corresponding PL mapping. (c) Raman and (d) PL spectra collected from the MoS₂ channel before and after the device transfer. The scale bar is 10 μm.*

The optical characterization of as-fabricated and transferred devices are studied using Raman and PL spectroscopy, which are widely used to obtain structural related information such as number of layers, strain, doping etc.[40] Figure 2 (a) represents an optical image of a device where monolayer MoS₂ channel with Ag electrodes are transferred to a pre-cleaned SiO₂/Si⁺⁺ substrate. A spatial mapping of spectrally integrated PL counts is initially performed on the transferred devices as shown in Fig. 2 (b). The bright and uniform luminescence from the monolayer MoS₂ indicates high quality optical response after DDT transfer. In a previous article, Jain *et. al.*[41] demonstrated dark spots and lines in PL mapping of PDMS-based transferred MoS₂ flakes indicating the interfacial bubbles and wrinkles, respectively. However, in our case, no such spots/lines are noticed which validates a clean and damage-free surface after hot-acetone treatment during DDT technique. To study the effects on the phonon modes of MoS₂, we have performed the Raman characteristics with the 532 nm excitation wavelength inside the channel area of the as-fabricated and transferred devices, as shown in Fig. 2 (c). The two prominent Raman active modes i.e. $E_{2g}^1$ and $A_{1g}$ are obtained in the high frequency region that corresponds to the in-plane and out-of-plane vibrations of constituent atoms and the frequency difference ($\Delta\omega$) between them is the characteristic signatures of number of layers/thickness of MoS₂.[34] Before transfer, the obtained $\Delta\omega$ is 20.3 cm⁻¹ which decreases to 19.3 cm⁻¹ as a result of the blue shift of $E_{2g}^1$ mode after following DDT method due to strain relaxation of monolayer MoS₂. However, it may be noted that no change or frequency shift is observed for $A_{1g}$ mode indicating absence of any charged interfacial impurities during transfer process. It may be noted that, the interaction of monolayers to the growth substrate is likely to be stronger than the transferred layer where a similar decrease of $\Delta\omega$ is observed in PDMS-based transferred films.[42] In addition to interfacial interaction with substrate, the presence of additional water soluble layer on the growth substrate may also cause the red shift of monolayer $MoS_2$. Therefore, the growth substrate induces an in-plane strain to the monolayer $MoS_2$, which is also validated elsewhere.[34] The corresponding strain is released after the transfer process causing a blue shift of $E_{2g}^1$ mode. Apart from frequency shifts, the full-width half maxima (FWHM) are calculated for both cases which remain almost unchanged for $A_{1g}$ modes (~ 4.8 cm$^{-1}$) indicating our transfer process doesn't compromise with the crystalline quality of the channel materials. However, FWHM of $E_{2g}^1$ modes are decreased from 3.5 to 2.7 cm$^{-1}$ which is also a signature of strain relaxation after transfer, as described earlier.[34] Furthermore, room temperature PL measurements are performed on $MoS_2$ channels and the results are shown in Fig. 2 (d). It is observed that before transfer, the PL spectra of $MoS_2$ are characterized by a strong excitonic transition at 1.84 eV (A exciton) and a relatively weak excitonic peak at 1.98 eV (B exciton), arising from the spin-orbit coupling induced valence band splitting. However, after transfer, a blue shift of 0.02 eV is observed for A excitons. The shift can be inferred in two possible ways, however difficult to differentiate at this stage. We expect the release of additional electron doping and strain from the water soluble layer on the growth substrate may cause a blue shift and the decrease of intensity can be attributed to the quenching of charged excitons as compared to neutral excitons after the transfer.[43] The results from Raman and PL spectra illustrates the retention of high quality films with intrinsic vibrational and optical properties after the DDT technique.

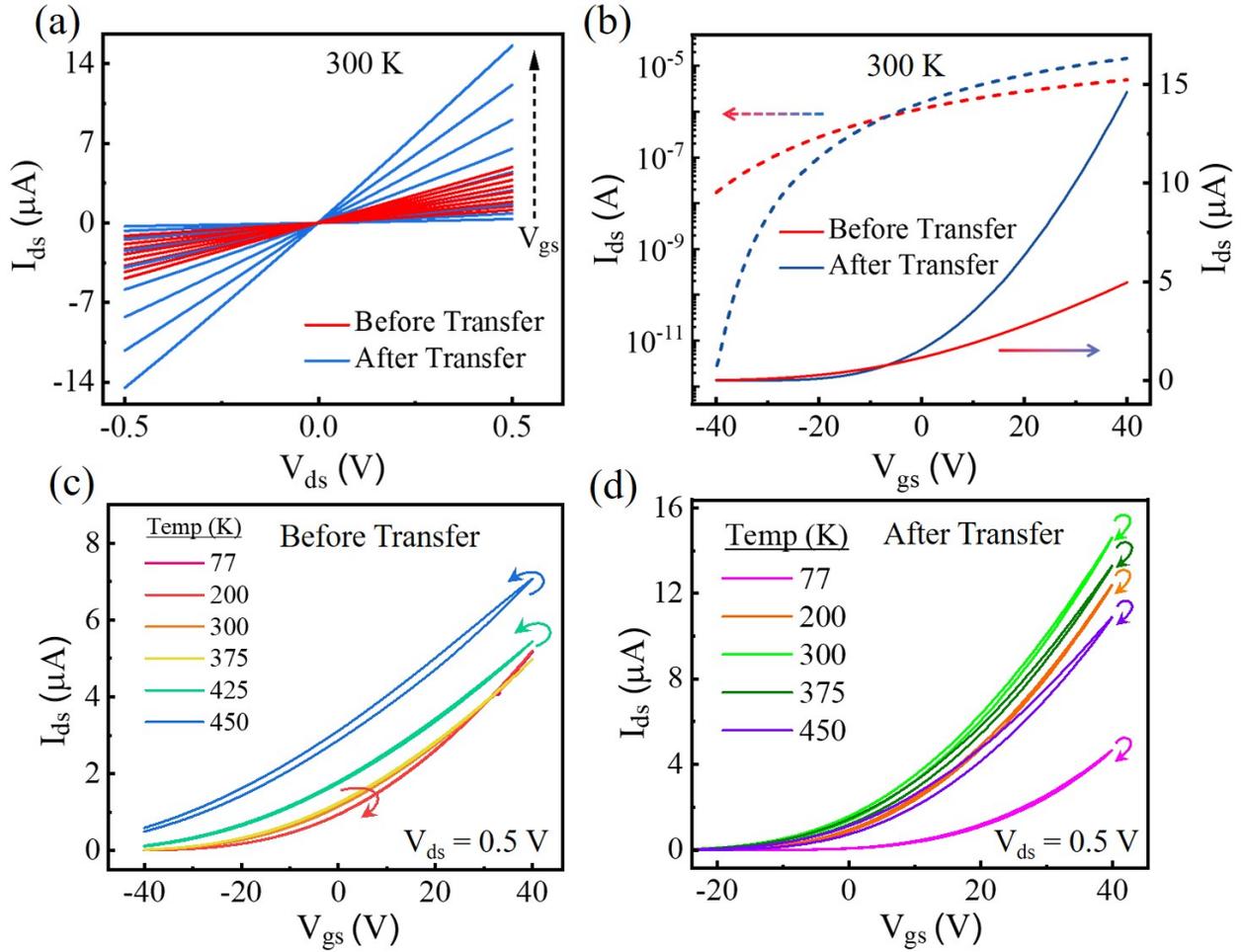

**Figure 3** *Room temperature (a) output and (b) forward sweep linear (left Y-axis) and semi-logarithmic (right Y-axis) transfer characteristics of a monolayer MoS$_2$ FET before and after the device transfer. V$_{gs}$ is varied from 0 to 40 V with step of 5 V in (a). Temperature dependence (ranging from 77 to 450 K) of dual sweep transfer curves of corresponding device (c) before and (d) after DDT method. The curved arrows represent the clockwise or anti-clockwise nature of the hysteresis loop.*

To examine the electrical properties of before and after the transfer of MoS$_2$ channels with source/drain electrodes, first we perform output characteristics, i.e. drain current (I$_{ds}$) vs drain-source voltage (V$_{ds}$) for gate voltages ranging from 0 to 40 V which are shown in Fig. 3 (a). The heavily doped Si$^{++}$ are treated as global back gate and the 300 nm SiO$_2$ generate required gate capacitance. The channel length and width of the devices are calculated to be 3 and 10 $\mu m$, respectively. The linear output characteristics are observed indicating the perseverance of Ohmic contact between Ag and MoS$_2$ for both before and after transfer cases. Such ohmic nature of MoS$_2$ based FETs using Ag electrodes are also reported in previous articles.[13,15,32] Moreover, we notice an enhanced gate tuning of drain currents in transferred devices. As shown in Fig. 1 (a), the large field-effect modulation from the gate voltage variations after the transfer indicate the absence of any pinning effects due to additional electron doping which is the case for as-fabricated device on growth substrate. The transfer characteristics, i.e. $I_{ds}$ vs gate-source voltage ($V_{gs}$), shown in Fig. 4 (b) indicates similar drain current enhancement while characterizing the device after the transfer. With the gate voltage sweep from -40 to 40 V, and a fixed $V_{ds}$ at 0.5 V, we obtain a low transistor on/off ratio about $10^2$ (before transfer) which increases to $> 10^6$ after the device transfer. Recently, Liu et. al.[32] also reported high on/off ratios ($>10^6$) in MoS$_2$ back-gated FET array with transferred Ag contacts which also validates the improved switching ratio after the transfer.

To investigate the carrier dynamics and trapping, the dual sweep transfer measurements are performed at various temperatures ranging from 77 to 450 K. The hysteresis which is inevitable in MoS$_2$ FETs owing to their large channel surface to volume ratio and charge trapping/detrapping at the channel/dielectric interfaces[13] are studied for both cases. As shown in Fig. 3 (c), before transfer, we observe a negligible clockwise hysteresis and minimal variation of transfer curves with increasing temperature up to 400 K. Interestingly, beyond 400 K we observe a hysteresis inversion and significant enhancement of hysteresis window with anti-clockwise behaviour. This anomaly in device performances could provide interesting results,[15] however may arise due to additional doping of charge carriers at high temperatures. Nevertheless, while characterizing the temperature dependent of dual sweep transfer curves after the device transfer, clockwise hysteresis are observed for all temperature range, as shown in Fig. 3 (d). In the next section, we will discuss more with few extracted transistor parameters for both cases.

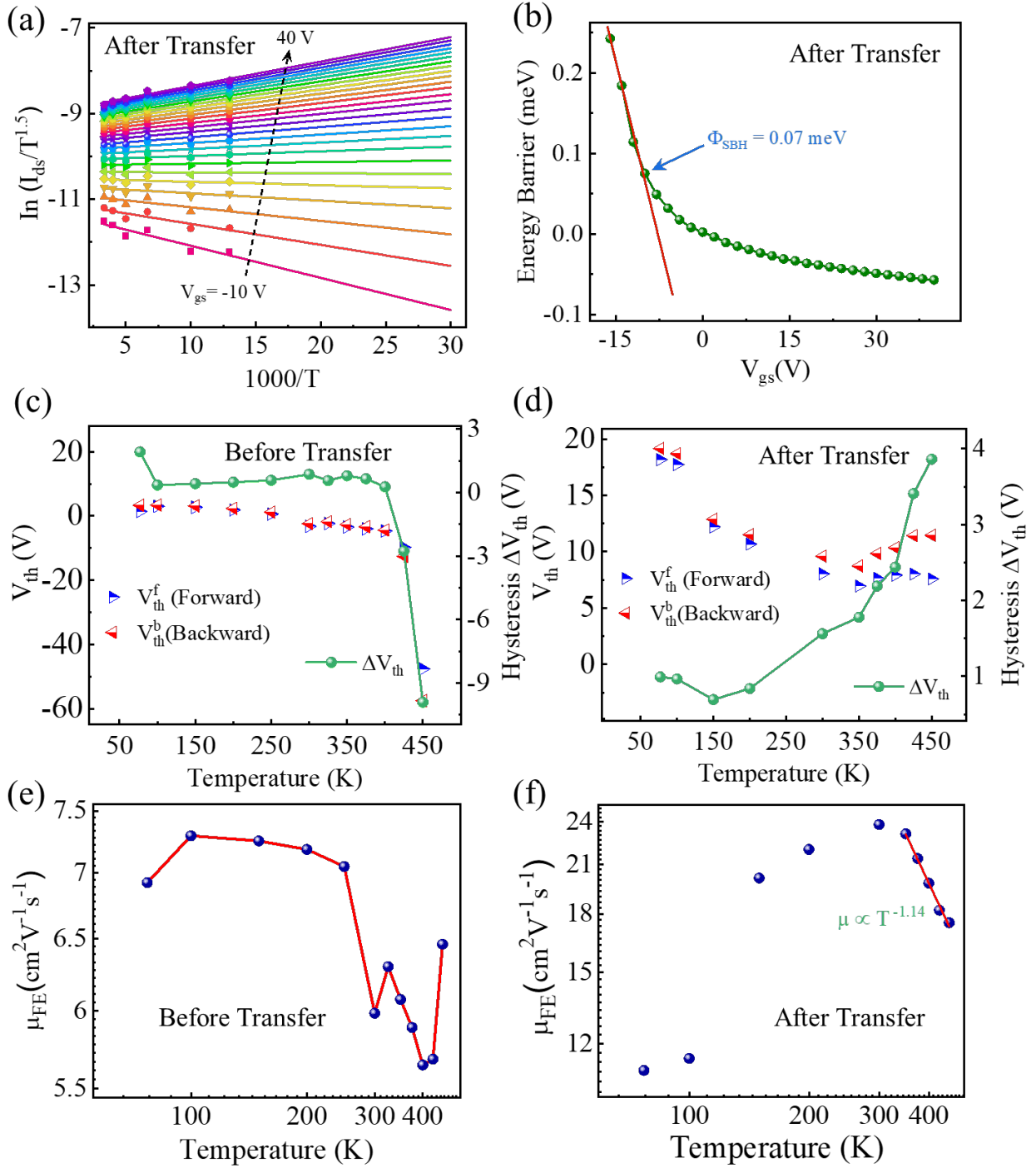

**Figure 4** *(a) Arrhenius plots for extraction of barrier heights (b) Barrier heights vs $V_{gs}$ and extraction of SBH from the linear fitting (red line) for Ag/MoS$_2$ contacts after device transfer.*

*Temperature dependence of threshold voltage shifts, $V_{th}^f$, $V_{th}^b$, (left Y-axis) and hysteresis window, $\Delta V_{th}$ (right Y-axis) of (c) before and after (d) device transfer. Temperature dependence of Field-effect mobility ($\mu_{FE}$) variations (e) before and (f) after the transfer. A linear fitting (red line) of the decrease of mobilities are shown at high temperatures ranging from 350 to 450 K in (f). The solid lines in (c, d, and e) are guide to the eye.*

Here, we want to emphasize that our proposed transfer technique demonstrates delamination of both channel and electrodes, which may severely affect the contact region (if not properly done or lack of mechanical support). To investigate more into the contact and hence current injection, the Schottky barrier height (SBH) between Ag and MoS$_2$ are extracted from the low temperature transfer characteristics described in the previous section. The SBH can be calculated from the following 2D thermionic emission equation[44]

$$I = A_{2D} W T^{\frac{3}{2}} \exp\left(-q\frac{\varphi_B}{K_B T}\right)\left[1 - \exp\left(-q\frac{V_{ds}}{K_B T}\right)\right] \quad (1)$$

Where $A_{2D}$ is the Richardson's constant for 2D systems[45], W is the channel width, q is the elementary charge, $K_B$ is the Boltzmann's constant, and $\varphi_B$ is the barrier height. Using this expression, we construct the Arrhenius plot as shown in the Fig. 4 (a) from which barrier heights as a function of V$_{gs}$ are extracted from their slopes. As the SBH corresponds to the barrier height at flat band condition that can be identified at the point where the extracted barrier heights deviates from its linear dependence.[44] As shown in Fig. 4 (b), the extracted SBH of the MoS$_2$ device after the transfer is approximately 0.07 meV, which is comparable to the SBH value of -0.02 meV in case of before transferred devices. The near equal SBH values further support robust and damage-free transfer of the devices.

In order to further investigate the carrier dynamics, the dependence of the threshold voltage shifts, evolution of hysteresis window ($\Delta V_{th}$) and calculated field-effect mobilities are plotted against the temperature for before and after device transfer, as shown in Fig. 4 (c-f). In our case, $\Delta V_{th}$ is defined by the threshold voltage difference between forward ($V_{th}^f$) and backward ($V_{th}^b$) sweeps. Before transfer, in Fig. 4 (c) the $\Delta V_{th}$ changes it's sign with the evolution of temperature, which is a clear signature of hysteresis inversion beyond 400 K. This behaviour may be attributed to additional ion doping arises from the growth substrates at high temperatures as demonstrated elsewhere.[15] Whereas, as shown in Fig. 4 (d), $\Delta V_{th}$ keeps on increasing with increasing temperature after the device transfer due to the thermally activated deep trapping states with enhanced carrier trapping and detrapping during forward and backward sweeps.[15] The temperature dependence of the $\mu_{FE}$ can be calculated by using the following equation[13]

$$\mu_{FE} = \frac{dI_{ds}}{dV_{gs}} \left[\frac{L}{WC_{ox}V_{ds}}\right] \qquad (2)$$

Where, L and W are the channel lengths and widths, and $C_{ox}$ being the capacitance of gate dielectric. The extracted mobilities show a nontrivial variation with temperature for as-fabricated device shown in Fig. 4 (e). However, after the transfer, mobility values are enhanced by a factor of 3 and the temperature dependence shows a similar variation as reported in previous articles.[15,46] As shown in Fig. 4 (f), the phonon limited carrier mobility is also observed to be proportional to $T^{-1.14}$, which is comparable with previous reports.[46] The results from the hysteresis windows and mobility variations with temperature indicate the proposed etchant-free device transfer technique could offer a promising avenue to achieve high performance electronic applications. The DDT method can be utilized to directly transfer the array of devices of 2D materials and heterostructures skipping various cumbersome steps in between. The proposed technique could be highly beneficial towards efficient FEOL and BEOL integrations of 2D materials with existing CMOS technology.

## Conclusion

In summary, we have demonstrated one-step PMMA-assisted etchant-free technique to directly transfer the devices with monolayer $MoS_2$ as channel materials and Ag as source/drain electrodes. The transfer technique results in high quality residue-free monolayer channels on a target $SiO_2/Si^{++}$ substrate as resulted from the spatial PL mapping. The Raman and PL spectra confirms the release of strain and removal of additional electron doping originated from the growth substate. Furthermore, reduced pinning effect is realized and the on/off ratio is increased by $10^4$ while characterizing the devices after transfer. The Schottky barrier height is found to be similar for both the cases illustrating the robust, damage free metal/2D semiconductor contact transfer using our technique. Moreover, the high temperature transport measurements provide insights into the carrier dynamics and trapping from the dual sweep hysteresis characteristics. The extracted mobilities are found to be nearly twice in the devices after the transfer. The temperature dependence of the threshold voltage and mobility variations validate the intrinsic transistor properties after the DDT method. Our work on the proposed transfer technique provides a facile approach towards efficient stacking and complex engineering of 2D material based future flexible devices.

## Acknowledgements

S.P.S. and R.P. thank to the Science and Engineering Research Board for partial financial support (CRG/2020/006190). We thank Gopal K Pradhan, KIIT Deemed to be University, Bhubaneswar for his help in Raman and PL measurements. We also acknowledge micro-Raman facility at the central research facility (CRF) of KIIT Deemed to be University, Bhubaneswar.

# Conflict of Interest

The authors declare no competing financial interest.